\documentclass{ws-procs975x65}

\begin{document}

\title{Dynamical Localization of Random Quantum Walks on the Lattice
\footnotetext{\copyright 2012 by the author}
}

\author{A. Joye}

\address{UJF-Grenoble 1, CNRS Institut Fourier UMR 5582,  Grenoble, 38402, France 
\\ E-mail: alain.joye@ujf-grenoble.fr
}

\begin{abstract}
This note describes recent results on the localization properties of Random Quantum Walks on the $d-$dimensional lattice in a regime analogous to the large disorder regime by means of the Fractional Moments Method adapted to the unitary framework.
\end{abstract}


\bodymatter

\section{Introduction}\label{intro}

The denomination Quantum Walks (QW for short) covers several variants of the definition we provide below \cite{ADZ, M, Ke, V-A} . Informally, a QW describes the discrete time quantum dynamics of a particle with internal degree of freedom, the quantum walker, on a lattice. This dynamics consists in making the walker jump between neighboring sites of the lattice. The Hilbert space of the particle is the tensor product of the finite dimensional Hilbert space of the internal degree of freedom, often called coin state in this context, with $l^2({\mathbb Z}^d)$, ${\mathbb Z}^d$ being the $d-$dimensional lattice. To meet the walk requirement, the one time step unitary operator $U$ allows transitions between sites of the lattice that are a finite distance away from each other only. The evolution at time $n\in \mathbb Z$ is then given by $U^n$. The dynamics of QW differs essentially from that generated by a Hamiltonian at integer times $n$ in that the latter usually allows for non-zero transitions between all sites, whereas the former forbids transitions between sites separated by more than a distance of order $|n|$.

Without attempting to be exhaustive, the variants alluded to concern, for example, the underlying configuration space which can be replaced by more general graphs \cite{Ke, AAKV} , the unitary framework which can be extended to completely positive maps\cite{AAKV, APSS} , the stationarity assumption which can be relaxed to accommodate time dependent walks\cite{AVWW, J3, HJ} , or the deterministic setup which can be enlarged to accommodate evolution operators taken randomly from a  set of unitary operators\cite{KLMW, Ko1}. The latter are called Random Quantum Walks (or RQW) and they describe the motion of a quantum walker in a static random environment. In such circumstances, it is expected that randomness induces destructive interferences that may lead in certain regimes to complete  suppression of transport due to  Anderson localization  \cite{Ki} . The results we describe here concern RQW of a certain type for which Anderson localization can be proven in certain regimes \cite{J4} , making use of previous works on certain types of random unitary operators, in particular \cite{HJS2, ABJ2} .


We give below a few topics and references in which QW play a role. The reader should consult the recent review  \cite{V-A} on QW for more informations, references and details.
QW provide discrete time models that can be used to describe the dynamics of certain quantum systems in appropriate regimes. It was demonstrated experimentally by \cite{Ketal, Zetal} that the one-dimensional dynamics of atoms trapped in certain time dependent optical lattices was accurately reproduced by a simple QW for times up to twenty iterations. Popular models in computational physics that go under the name quantum networks also belong to the class of RQW. A distinguished example is the so called Chalker Coddington model \cite{CC} of condensed matter physics which describes the approximate dynamics of two-dimensional electrons in a random background potential subject to a large perpendicular magnetic field. The quantum computing community is interested in QW for quite different reasons. Namely, due to their simple algorithmic implementation, QW play an important role in assessing the efficiency of elaborated quantum algorithms to be used on quantum computers, in the same way classical random walks are used in theoretical computing, see {\it e.g.} \cite{S}. Finally, due to the probabilistic interpretation of quantum mechanics, certain versions of QW have been viewed by probabilists as quantum generalizations of classical random walks on the lattice, with quite different properties, see \cite{Ko} . 

\section{Setup}

The type of QW we address here is one of the simplest instances of QW called Coined Quantum Walk. The one time step evolution operator is defined by the action of a unitary coin matrix $C$ on the internal degree of freedom only, followed by a one step shift on the lattice, conditioned on the state of the coin. We shall be interested in the case where the coin matrix depends on the position of the walker in a random fashion, so that Anderson localization can take place for the resulting RQW in some regimes we analyze.


Let us describe precisely the mathematical framework and notations we use. The Hilbert space of the walker is 
\begin{equation}
{\cal H}={\mathbb C}^{2d}\otimes l^2({\mathbb Z}^d), \ \ \ \ d\geq 1.
\end{equation}
The canonical basis of the coin state space is denoted by $\{|\tau \rangle\}_{\tau \in I_\pm}$, where $I_\pm=\{\pm 1, \dots, \pm d\}$ and that of $l^2({\mathbb Z}^d)$
by $\{|x\rangle\}_{x\in {\mathbb Z}^d}$. Consequently, we denote the basis of ${\cal H}$ obtained by tensor products by  $\{|\tau, x\rangle\}_{\tau \in I_\pm, x\in {\mathbb Z}^d}$, where $|\tau, x\rangle=|\tau\rangle\otimes|x\rangle$.

We start by defining a reference Coined Quantum Walk. We need a jump function  $r:I_\pm\rightarrow {\mathbb Z}^d$ given by $r(\tau)=\mbox{sign} (\tau)e_{|\tau|}$, where $\{e_j\}_{j=1,\dots, d}$ denotes the canonical basis of ${\mathbb R}^d$. Then, the coin state dependent shift operator is defined by 
\begin{equation}
{S}:=\sum_{x\in{\mathbb Z}^d}\sum_{\tau\in I_\pm}P_\tau\otimes |x+r(\tau)\rangle\langle x|,
\end{equation} 
where $P_\tau$ is the orthogonal projector on the state $|\tau\rangle$. For  $C\in U(2d)$ a unitary coin matrix, the corresponding Coined Quantum Walk is defined by
\begin{equation}\label{cqw}
U(C)=S (C\otimes {\mathbb I})=\sum_{x\in{\mathbb Z}^d}\sum_{\tau\in I_\pm}P_\tau C\otimes |x+r(\tau)\rangle\langle x|.
\end{equation}
The coin matrix $C$ is considered as a parameter of the QW, hence the notation $U(C)$. Note that the definition (\ref{cqw}) is translation invariant, so that, for generic matrices $C$, the operator $U(C)$ is absolutely continuous and induces ballistic motion, by a RAGE type argument. However, certain choices of $C$'s give rise to QW that do not propagate, as we will see below.

\subsection{Random Quantum Walk}

The environment of the walker is made random by decorating the elements of the coin matrix $C$ by phases which are site-dependent and random as follows. Set 
\begin{equation}
C_{\omega}({x})_{\tau,\sigma}=\exp({i{\omega}_{{x}+r(\tau)}^\tau})C_{\tau,\sigma}, \ \ \ \forall \ \ \sigma, \tau \in I_\pm.
\end{equation} 
The corresponding random one time step evolution operator is then defined by
\begin{equation}\label{sitedep}
U_\omega(C)=\sum_{x\in{\mathbb Z}^d}\sum_{\tau\in I_\pm}P_\tau C_\omega(x)\otimes |x+r(\tau)\rangle\langle x|.
\end{equation} 
Introducing the unitary diagonal random operator ${\mathbb D}_\omega=\mbox{diag}(e^{i\omega^\tau_x})$ with respect to the basis $\{|\tau, x\rangle\}_{(\tau, x)\in I_\pm\times {\mathbb Z}^d}$
we have the relation
\begin{equation}\label{struct}
U_\omega(C)={\mathbb D}_\omega U(C).
\end{equation} 
We will assume that $\{ \omega_{x}^{\tau} \}^{\tau\in I_\pm}_{x\in{\mathbb Z^d}}$ are i.i.d. ${\mathbb T}$--valued random variables distributed according to
$d\mu(\theta)=l(\theta)d\theta$  with   $0\leq l\in L^\infty $. The corresponding probability space is denoted by $(\Omega=\mathbb T^{{\mathbb Z}^d\times I_\pm}, \mathbb P=\otimes_{\tau,x}d\mu )$ and a realization is denoted by $\omega\in \Omega$. This provides us with an ergodic unitary operator $U_\omega(C)$ defining our RQW.


Next, we argue according to \cite{J4} that dynamical localization takes place for $U_\omega(C)$, provided $C$ is close enough to a coin matrix that forbids propagation and induces a fully localized walk.

\subsection{Fully Localized Walk}

We consider here a case of coin matrix which forbids any propagation. 
Let $\pi\in{\mathfrak S}_{2d}$ be a permutation acting on $I_\pm$ 
and set
\begin{equation}
{C_\pi}=\sum_{\tau\in I_\pm}| \pi(\tau)\rangle\langle \tau|. 
\end{equation}
In case  $\pi$ has the form $\pi=(\tau, \pi(\tau), \dots, \pi^{2d-1}(\tau))$, one checks that the $U_\omega(C_\pi)$-cyclic subspaces generated by any basis vector $|\tau,x\rangle$ are given by
\begin{equation}\label{cyclic}
{{\cal H}^\tau_x}=\mbox{span}\Big\{|\tau,x\rangle, |\pi(\tau),x+r(\pi(\tau))\rangle, \dots, \big|\pi^{2d-1}(\tau),x+\sum_{s=0}^{2d-1}r(\pi^s(\tau))\big\rangle\Big\} 
\end{equation}
and that their sum yields ${\cal H}$. This shows in particular that $U_\omega(C_\pi)$ is pure point, for any realization $\omega\in\Omega$, and that for any initial condition $\psi\in\cal H$ with finite support on $\mathbb Z^s$, $U^n_\omega(C_\pi)\psi$ has finite support as well, uniformly in $n\in\mathbb Z$. Moreover, the spectrum of the restriction of $U_\omega(C_\pi)$ to ${\cal H}^\tau_x$ is given by
\begin{equation}\label{specloc}
\sigma(U_\omega(C_\pi)|_{{\cal H}^\tau_x})=e^{i\alpha_x^\tau(\omega)/2d}\{1, e^{i2\pi/2d}, \cdots, e^{i2\pi(2d-1)/2d}\},
\end{equation}
where 
\begin{equation}
\alpha_x^\tau(\omega)=\omega_x^\tau+\omega_{x+r(\pi(\tau))}^{\pi(\tau)}+\dots+\omega_{x+\sum_{s=0}^{2d-1}r(\pi^s(\tau))}^{\pi^{2d-1}(\tau)}
\end{equation}
are i.i.d. random phases distributed according to the $2d$-fold convolution of $d\mu$. Consequently, we have access to all probabilistic properties of the eigenvalues of $U_\omega(C_\pi)|_{{\cal H}^\tau_x}$.

\section{Main Result}

We now have all the ingredients to state the dynamical localization result of  \cite{J4} , under the hypotheses made so far. Throughout, we use the norm $|x|=\max_{i=1,\dots,d}|x_i|$ on $\mathbb Z^d$ and we use the notation ${\mathbb U}=\{|z|=1\}$.

\begin{theorem} \label{thdynloc}
There exists  $\delta>0\ K<\infty$, ${\gamma}>0$ such that $\forall \ C\in U(2d)$, $\|C-C_\pi\|_{\mathbb C^d}<\delta$ implies $\forall \ {x, y} \in \mathbb Z^d$ and $\forall  \sigma, \tau \in I_\pm$
\begin{equation}\label{dynloc}
{\mathbb  E} \left[ {\sup_{{ f}\in C({\mathbb U}),\, \|{ f}\|_\infty\leq 1}}\  |\langle {\tau }, {x} |, { f}(U_{ \omega}(C)) \, |{ \sigma}, {y} \rangle| \right]\leq Ke^{-{\gamma } {|x-y|}}.
\end{equation}
\end{theorem}
As in the self-adjoint case, dynamical localization implies almost surely finiteness of all the moments of the position operator $X$ (defined by $X |\tau,x\rangle=x|\tau, x\rangle$), as well as spectral localization 
\begin{corollary} If Theorem \ref{thdynloc} holds, there exists $\Omega_0\subset\Omega$ with ${\mathbb P}(\Omega_0)=1$ such that if $\psi_0$ has finite support, $p\in\mathbb N$ and $\omega\in \Omega_0$,
\begin{equation}
\sup_{{n}\in{\mathbb Z}}\| |X|^{p}U^{{n}}_{{\omega}}(C){\psi_0}\|\leq K_{{\omega}}<\infty, 
\end{equation}
and  $\sigma(U_{ \omega}(C))$ is pure point.
\end{corollary}
\begin{remark}
We interpret the condition that $C$ be close to $C_\pi$ as an analog of the large disorder regime in the Anderson model. The heuristics being that for large disorder, the Anderson Hamiltonian is dominated by the potential which does not induce transitions on the lattice, a feature shared by $U_{{\omega}}(C_\pi)$.  \\
In keeping with the self-adjoint case, if $d=1$, dynamical localization is shown in \cite{JM} to take place for all values of the parameter $0<r<1$ for $C=\begin{pmatrix}t&r\cr r&-t\end{pmatrix}$, and not only for $r$ close to one. More general one-dimensional RQW for which dynamical localization holds have since been studied, see \cite{SK, ASW} . \\
RQW were first introduced in \cite{Ko1} with a special choice of random phases which gives rise to ballistic transport.
\end{remark}

\section{Some Steps of the Proof}

We use an adaptation to the unitary framework of the Fractional Moments Method of Aizenman-Molchanov \cite{AM} to prove localization provided in \cite{HJS2} . The Multi-Scale Analysis  approach introduced by Fr\"ohlich-Spencer \cite{FS} can most probably be adapted to the present context as well.  

\subsection{Fractional Moments Method}

Dynamical localization (\ref{dynloc}) for random unitary operators $U_\omega(C)$ of the form  (\ref{struct}) is shown in \cite{HJS2} to follow once one proves an estimate on the Green function 
\begin{equation}
G_{\tau,\sigma,\omega}(x,y;{z})=\langle \tau,x | \, (U_\omega(C)-z)^{-1} | \sigma,y\rangle
\end{equation} 
of the form:  \medskip \\
$ \exists \ 0<s<1$, $\gamma(s)>0$, and   $K(s)<\infty$ such that $\forall \ z\not\in\mathbb U$ and $\tau, \sigma\in I_\pm$,
\begin{equation} \label{AM}
{\mathbb E}(|G_{\tau,\sigma,\omega}(x,y;{z})|^{s})\leq K({s})e^{-\gamma({s}) |x-y|}.
\end{equation}

We provide the main general arguments of  \cite{HJS2} to back this statement in the Appendix and admit this criteria for now. In the following, the symbol $c$ will denote a generic unimportant constant, which may change from line to line. 


In order to prove (\ref{AM}) for the case at hand, we resort to finite volume estimates, adapting \cite{AENSS} to the unitary case along the lines of \cite{HJS2} . This is required because $U_\omega(C)$ is off-diagonal, which forbids the use of the direct infinite volume approach developed in \cite{J} . Finite volume estimates were used successfully in the study \cite{ABJ2} of the Chalker-Coddington model.

The general strategy consists in defining boundary conditions compatible with unitarity in order to define restrictions of $U_\omega(C)$ to Hilbert spaces ${\cal H}^{\Lambda}$ associated with cubes $\Lambda=\{x\in \mathbb Z^d, |x|\leq L\}$ of side length $2L+1\in\mathbb N$. By ergodicity, it is enough to considered cubes centered at the origin. Then one establishes estimates on the LHS of (\ref{AM}) for the associated finite volume Green function. This is done using perturbation theory in $C-C_\pi$ and the properties of  (the restriction to ${\cal H}^{\Lambda}$ of) the localized walk $U_\omega(C_\pi)$. Finally, the link between finite and infinite volume estimates on the Green function is made along the lines of \cite{HJS2} , via geometric resolvent identities, decoupling estimates and iterations. 
We only describe below the construction of the unitary restrictions and the finite volume fractional moments estimates which are model dependent. We refer the reader to \cite{HJS2} for the last step of this strategy, which is quite general and more involved.

\subsection{Finite Volume Restrictions}

We first introduce boundary conditions. Consider the QW described by 
\begin{equation}\label{bc}
U^L(C)=\sum_{x\in{\mathbb Z}^d}\sum_{\tau\in I_\pm}P_\tau C(x)\otimes |x+r(\tau)\rangle\langle x|,
\end{equation} 
where $C(x)=C_\pi$ if $|x|\in\{L-1, L, L+1\}$ and $C(x)=C$ otherwise. By construction, if the walker sits on a site $x$ such that $|x|=L$, its environment consists in matrices $C_\pi$ which forces it to remain in the corresponding subspaces ${\cal H}_x=\oplus_{\tau\in I_\pm}{\cal H}^\tau_x$, see (\ref{cyclic}). This creates an impenetrable  boundary. Therefore, introducing 
\begin{equation}
{\cal H}^\Lambda=\oplus_{|x|\leq L}{\cal H}_x\ \ \mbox{and}\ \ {\cal H}^{\Lambda^C}\ \  \mbox{by}\  \ {\cal H}={\cal H}^\Lambda\oplus{\cal H}^{\Lambda^C},
\end{equation}
 we have ${U^L(C)} {\cal H}^\Lambda\subset {\cal H}^\Lambda$ and ${U^L(C)} {\cal H}^{\Lambda^C}\subset {\cal H}^{\Lambda^C}$. Also, the randomized version 
\begin{equation}
U^L_\omega(C)={\mathbb D}_\omega U(C)^L \  \ \mbox{satisfies} \ \ {U_\omega^L(C)} {\cal H}^\Lambda\subset {\cal H}^\Lambda, \ \  {U_\omega^L(C)} {\cal H}^{\Lambda^C}\subset {\cal H}^{\Lambda^C}
\end{equation}
and moreover, uniformly in $\omega$ and $L$,
\begin{equation}
\|U_\omega(C)-U_\omega^L(C)\|\leq c\|C-C_\pi\|_{\mathbb C^{2d}}.
\end{equation}
 We can now introduce the unitary finite volume restriction of the RQW by
 \begin{equation}
 U^{\Lambda}_\omega(C)=U_\omega^L(C)|_{{\cal H}^\Lambda},
 \end{equation}
with corresponding finite volume Green function $G_{\tau,\sigma,\omega}^{\Lambda}(x,y;z)$. Note that for  $C=C_\pi$, $U^\Lambda_\omega(C_\pi)=U_\omega(C_\pi)|_{{\cal H}^\Lambda}$ is a direct sum of $O(L^d)$ matrices of the form $U_\omega(C_\pi)|_{{\cal H}^\tau_x}$.

\subsection{Finite Volume Fractional Moment Estimates}

With the goal of eventually taking the limit $L\rightarrow \infty$, we focus on the LHS of (\ref{AM}) with $G^\Lambda$ in place of $G$. We first estimate the probability to find eigenvalues of $U^{\Lambda}_\omega(C)$ close to a point $z\not\in\mathbb U$, if $C$ is close to $C_\pi$. 
\begin{lemma} For any $z\not\in \mathbb U$,
$\exists \ c_0, c_1>0 $ such that if $\|C-C_\pi\|_{\mathbb C^{2d}}\leq c_0\eta$, with $\eta L^{d}$ small enough,
\begin{equation}
{\mathbb P}(\mbox{\em dist} \, (\sigma(U^{\Lambda}_\omega(C),z)> \eta)\geq 1-c_1\eta L^{d}.
\end{equation}
\end{lemma}
This estimate is based on 
${\mathbb P}(\sigma(U_\omega(C_\pi)|_{{\cal H}^\tau_x})\cap A =\emptyset)\geq 1-c|A|$, for a small arc $A\subset \mathbb U$ by independence of the random phases, see (\ref{specloc}),  and on perturbation theory. Together with the trivial estimate 
$|G^\Lambda_{\tau,\sigma,\omega}(x,y;{z})|\leq 1/\mbox{dist } (\sigma(U^{\Lambda}_\omega(C)),z)$, it yields
\begin{proposition}
For any $0<{s}<1$, ${\alpha}>0$, there exists $c >0, \beta >0$ such that ${\eta=L^{-\beta}}$ with  ${L}>\hspace{-.1cm}> 1$ implies  
\begin{equation}
{\mathbb E}(|G^\Lambda_{\tau,\sigma,\omega}(x,y;{z})|^{s})\leq c/L^{\alpha}\end{equation}
 for any $z\not\in \mathbb U$, $\tau,\sigma\in I_\pm$,  $x,y\in \Lambda $ such that $2\leq |x-y|$.
\end{proposition}
\begin{remark} A similar finite volume estimate was first shown to hold true for the Chalker-Coddington model in \cite{ABJ2} by an argument we adapted to the model at hand. From here on, thanks to relation (\ref{struct}), we can argue as in Section 13 of \cite{HJS2}  to reach the sought for estimate (\ref{AM}) , see \cite{J4} .
\end{remark}

\appendix{Fractional Moment Estimates Imply  Localization}

We provide here the key steps from \cite{HJS2} behind the statement "(\ref{AM}) implies (\ref{dynloc})". The statement is based on the specific form (\ref{struct}) of $U_\omega(C)$, on the fact that the deterministic part $U(C)$ has a band structure and on the assumption on the randomness of the phases in $\mathbb D_\omega$.

One first establishes the following preliminary estimate: \\
 $\forall \ 0<{s}<1$, $\exists \ K({s})<\infty$ such that 
\begin{equation} 
\int\int |G_{\tau,\sigma, \omega}(x,y; {z})|^{s} l(\theta_x^\tau) l(\theta_y^\sigma)\ d\theta_x^\theta d\theta_y^\sigma \leq K({s}),
\end{equation}
for all ${z}\not\in{\mathbb U}, \forall x, y\in {\mathbb Z}^d $. It implies 
${\mathbb E} (|G_{\tau,\sigma, \omega}(x,y;{z})|^{s}) \leq K({s})$. From there, one derives a "second moment estimate" {\it \`a la Graf}\cite{G} relating, roughly, the expectation of the square of the Green function to the fractional moments 
\begin{equation} \label{graf}
 {\mathbb E} (1-|{z}|^2)(|G_{\tau,\sigma, \omega}(x,y;{z})|^{2}) \leq K({s})
\sum_{{|m-x|\leq 4}}  \max_{(\tau',\sigma')\in I_\pm^2}{\mathbb E} (|G_{\tau',\sigma', \omega}(m,y;{z})|^{s}).
\end{equation}
This estimate is used in the functional calculus for unitary operators expressed as follows. For ${f}\in C({\mathbb U})$ and $U$ a unitary operator,
\begin{equation} 
f(U)=w-\lim_{r\rightarrow 1^-}\frac{1-r^2}{2\pi}\int_{0}^{2\pi}(U-re^{i\theta})^{-1}(U^{-1}-re^{-i\theta})^{-1}f(e^{i\theta})\, d\theta.
\end{equation} 
Note the factor $1-r^2=1-|re^{i\theta}|^2$ and the square of the resolvent $(U-re^{i\theta})^{-1}{(U-re^{i\theta})^{-1}}^*$ which correspond to the LHS of (\ref{graf}). Thus, putting everything together, one deduces that if (\ref{AM}) holds for $U_{{\omega}}(C)$, then there exist $0<K, \gamma <\infty$ such that $\forall \ x, y\in {\mathbb Z}^d $, $\forall \tau,\sigma \in I_\pm^2$, (\ref{dynloc}) holds:
\begin{equation} 
{\mathbb E}\,\Big(\sup_{{f}\in C({\mathbb U})\atop \| f\|_\infty\leq 1}|\langle \tau, x |{f}(U_{{\omega}}(C))|\sigma, y\rangle| \,\Big)\leq Ke^{-\gamma|x-y|}. 
\end{equation}

\section*{Acknowledgments} Work partially supported by the Agence Nationale de la Recherche, grant ANR-09-BLAN-0098-01

\end{document}